# QDNA-ID Quantum Device Native Authentication

Osamah N. Neamah

Department of Mechatronic Engineering, Graduate Institute, Karabuk University, Karabuk, Turkey.

2238173003@ogrenci.karabuk.edu.tr

***Abstract:*** QDNA-ID is a trust-chain framework that links physical quantum behavior to digitally verified records. The system first executes standard quantum circuits with random shot patterns across different devices to generate entropy profiles and measurement data that reveal device-specific behavior. A Bell or CHSH test is then used to confirm that correlations originate from genuine non classical processes rather than classical simulation. The verified outcomes are converted into statistical fingerprints using entropy, divergence, and bias features to characterize each device. These features and metadata for device, session, and random seed parameters are digitally signed and time stamped to ensure integrity and traceability. Authenticated artifacts are stored in a hierarchical index for reproducible retrieval and long-term auditing. A visualization and analytics interface monitors drift, policy enforcement, and device behavior logs. A machine learning engine tracks entropy drift, detects anomalies, and classifies devices based on evolving patterns. An external verification API supports independent recomputation of hashes, signatures, and CHSH evidence. QDNA-ID operates as a continuous feedback loop that maintains a persistent chain of trust for quantum computing environments.

***Keywords:*** Quantum Noise Characterization, Quantum Hardware Attestation, Device-Independent Verification, Quantum Provenance, Quantum System Integrity.

1. ## Introduction

In the last decade, quantum computing has witnessed significant and rapid development, moving from theoretical models to commercial cloud platforms. This development has led to fundamental limitations related to quantum device authenticity and result integrity. With the increasing reliance on open APIs (such as IBM-Q, IonQ, and Rigetti), it has become difficult to distinguish between a real quantum processor operation and one that has been simulated or modified in software [1]. Over time, the problem will begin to escalate in untrusted environments where quantum results can be falsified or reused without evidence of the physical source that generated them. Recent studies in recent years have addressed this issue from different angles. Some research has focused on noise fingerprints to characterize and recognize the physical properties of devices through statistical analysis of quantum circuit outputs [2], [3]. Others studies to went to develop classification algorithms based on state-level distributions to recognize devices while monitoring quantum drift. Most of these efforts have been limited to the analytical aspect and have not provided a cryptographic chain of trust linking quantum physical behavior to independently verifiable records [4], [5], [6].

The Quantum-based Noise-dependent Fingerprint (Q-ID) is a lightweight fingerprint that uses minimal quantum resources and measures the performance difference of a user's circuit under two levels of noise. This work focuses on identifying and authenticating quantum servers within a quantum network, which has been experimentally verified on the IBM quantum platform [7]. Researchers presented a practical large-scale localization algorithm using quantum fingerprints (QLoc). QLoc encodes wireless signal strength vectors into quantum states to compute similarities, and this method achieves significant improvements in spatial and temporal complexity compared to classical fingerprinting methods [8]. Learning from quantum device fingerprints using machine learning is presented in [9]. Machine learning is used to classify noise distributions over time from measured output probabilities. This work reveals that each quantum device exhibits a unique and evolving noise profile or "fingerprint" that can be learned and monitored. This paper presents QPP-RNG, a true random number generator (TRNG) for producing high-quality randomness that leverages the proposed methodology of system-level fluctuations, CPU timing, and dynamic RAM refresh disturbances. The researchers use a quantum phase plate (QPP) to amplify microscopic temporal fluctuations into strong entropy, and using the NIST SP 800-90B IID standard and related tests, it achieves superior minimum entropy (approximately 7.9 bits/byte), surpassing commercial random number generators [10].

In cryptography, researchers have presented post-quantum cryptography (PQC) methods to secure Internet of Things (IoT) systems against quantum threats [11], such as practical device-independent quantum cryptography via entropy accumulation. This theoretical work establishes a framework for device-independent quantum cryptography, meaning that security does not rely on trusted devices. To enable rigorous proofs for device-independent quantum key distribution, it utilizes entropy accumulation by defining the total entropy as the sum of individual components. Current loophole-free Bell tests make these device-independent protocols experimentally feasible, pushing towards achieving the highest levels of cryptographic security [12]. One of the most significant limitations of current quantum devices is noise from external environmental factors such as temperature, humidity, or chemical reactions within them over time, which requires periodic maintenance and calibration. Quantum device noise can be tracked over time, creating a unique fingerprint for the device, generating digital signatures, and verifying the integrity of each fingerprint file to ensure system provenance, which is the core of this proposed study (See Table 1.).

Table 1. Comparison of quantum fingerprinting and cryptography studies.

| Ref. | Core | Quantum Concept | Application | Contribution | Evaluation Platform |
|---|---|---|---|---|---|
| [7] | Authentication among quantum servers | Noise-based performance gap as fingerprint | Quantum network security | Efficient, low-resource quantum device ID | IBM Quantum Platform |
| [8] | Scalable indoor localization | Quantum fingerprinting, similarity measure | Localization / Positioning systems | Exponential speedup in time & space | IBM Quantum Cloud (testbed) |
| [9] | Characterizing device-specific | Noise fingerprinting via ML | Quantum device diagnostics | Identified evolving noise signatures | IBM Quantum Cloud |

| | quantum noise | | | using ML | |
| --- | --- | --- | --- | --- | --- |
| [10] | Random number generation | Quantum permutation pad (QPP) | Cryptography, IoT security | High-entropy TRNG from classical hardware jitter | Multi-platform |
| [11] | Post-quantum cryptographic methods | Post-quantum algorithms, hybrid systems | IoT authentication | Review of PQC challenges and directions | Conceptual / Analytical |
| [12] | Theoretical framework for DI cryptography | Entropy accumulation, Bell non-locality | Quantum key distribution | Security proof for DI-QKD with feasible parameters | Theoretical with Bell-test experimental base |

Based on the literature researches gap, the QDNA-ID project presents an integrated framework for what can be called "quantum provenance," a system that combines physical verification, statistical analysis, digital signatures, and authenticated timestamps within a single chain. The framework based on implementing the Bell/CHSH test to verify the non-locality of the device, followed by extracting a set of multi-layer statistical properties such as Shannon entropy, Jensen–Shannon divergence, Gini coefficient, perplexity, and $p_0$ bias to generate a unique quantum device fingerprint for each run. This fingerprint is encrypted in a file and accompanied by an HMAC signature and an RSA signature along with provenance metadata including processor, time, version, and calibration parameters. The results of this process are displayed via an interactive dashboard built with visualization, enables chronological fingerprint comparison, drift analysis, and change detection through different quantum devices. The system also offers a Public or Redacted Mode that enables researchers to share signed data without revealing sensitive details, that's achieves a balance between scientific transparency and cryptographic security. This QDNA-ID represents a qualitative step toward transforming quantum records from tamper-evident data into forensically-verified digital records (See Figure 1.). QDNA-ID introduces a new dimension to the concept of reproducibility in quantum science with laying the foundation for a future infrastructure that can support Web3 Quantum Provenance Networks and trusted research supply chains.

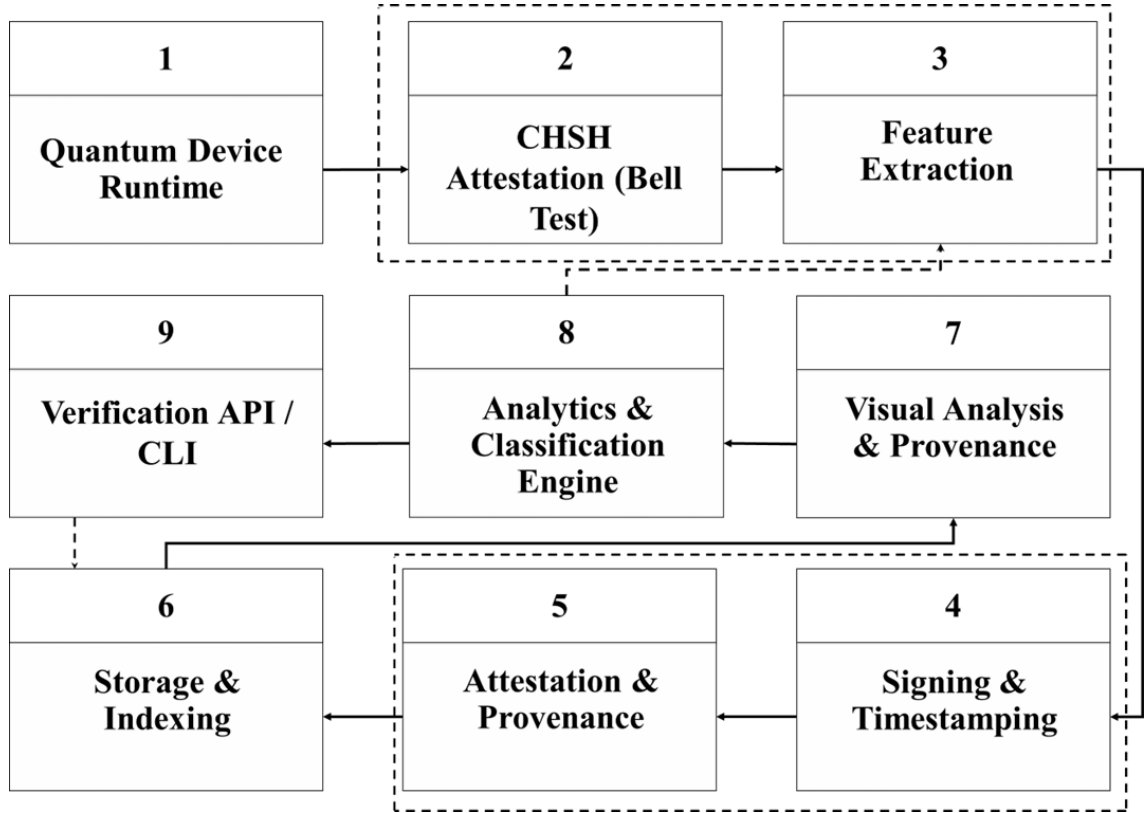

Figure 1. System-Level Architecture for Quantum Device Native Authentication (QDNA-ID).

## 2. Claim

1. A quantum execution subsystem configured to execute a plurality of quantum circuits on a physical quantum processor and to collect measurement outcomes representing the physical behavior of that device;
   - **Prior Art:** Quantum hardware routinely executes parameterized quantum circuits and collects measurement outcomes for calibration and benchmarking.
   - **Distinction**: Subsystem explicitly for device-behavior profiling (noise, coherence, entanglement signatures) as input to a provenance pipeline. This purpose-driven use of measurement outcomes for device identity is not taught in the cited works.
2. A quantum verification subsystem operative to perform one or more non-classical attestation procedures, including Bell-type or equivalent tests, to confirm that said measurement outcomes originate from genuine quantum processes;
   - **Prior Art:** Device-independent certification by Bell/CHSH tests is well established for verifying quantum behavior of devices.

- **Distinction:** Integration of such attestation into an operational hardware-identity and provenance system, producing a session-level "authenticity score" tied to fingerprinting.

3. A feature extraction and characterization module configured to derive statistical and information-theoretic descriptors from said outcomes to form a probabilistic fingerprint uniquely associated with the device;
    - **Prior Art:** Noise-fingerprinting of quantum devices has been demonstrated (e.g., IBM machines) with high classification accuracy.
    - **Distinction:** Use of information-theoretic descriptors (entropy, divergence, bias) to generate a probabilistic fingerprint linked to device identity and then cryptographically bound is beyond existing works.

4. A cryptographic provenance engine configured to generate a digitally signed and timestamped record linking said fingerprint and verification results, thereby producing a tamper-evident provenance artifact;
    - **Prior Art:** Digital-signature, timestamping, and tamper-evident logging exist broadly in IT and security domains.
    - **Distinction:** A provenance engine that binds together quantum-device fingerprint, authenticity score, and metadata into a single tamper-evident artifact for quantum hardware identity.

5. A validation and storage subsystem configured to verify said provenance artifacts using corresponding public credentials and to maintain an immutable or append-only record of verified fingerprints (See Figure 2).

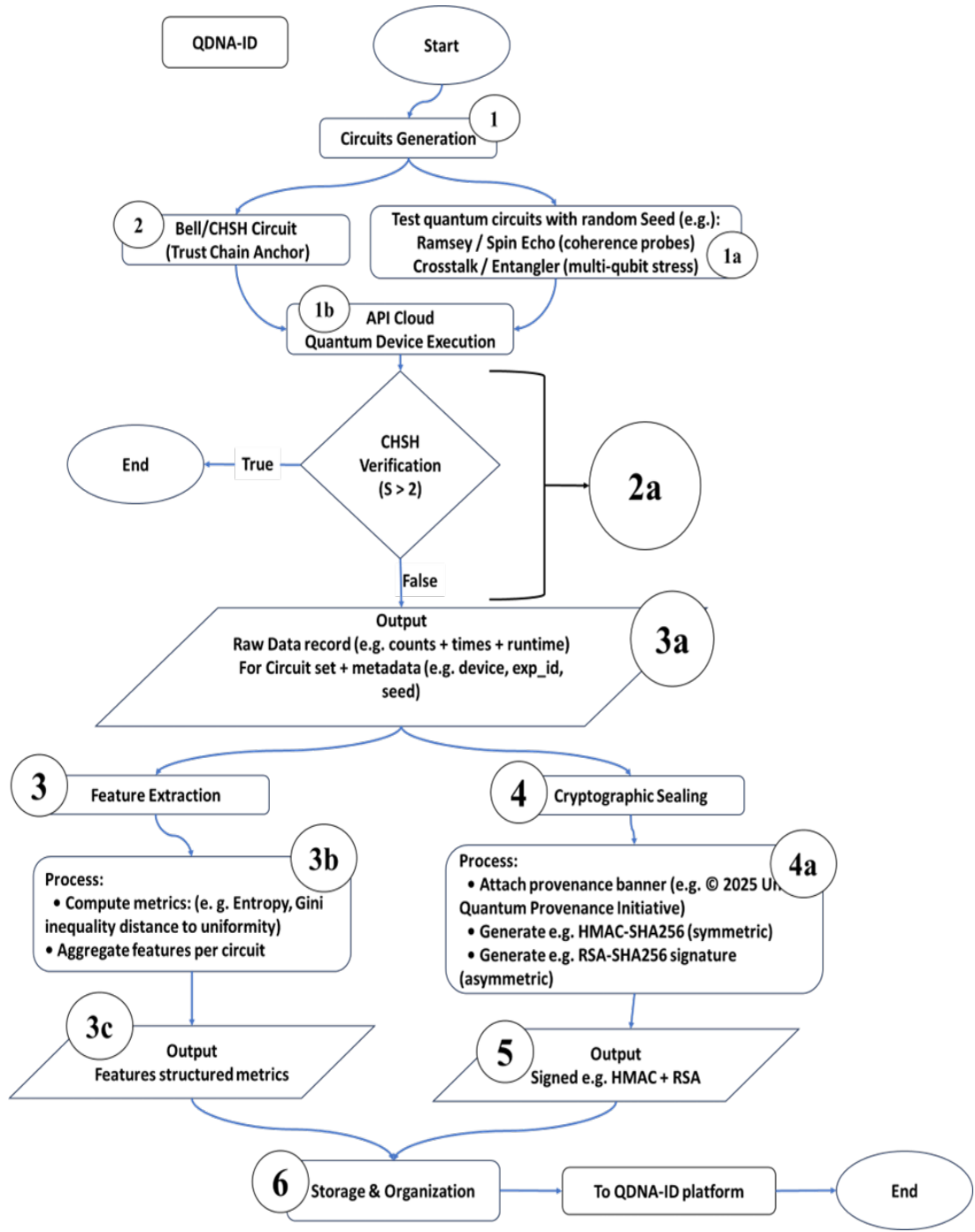

(a)

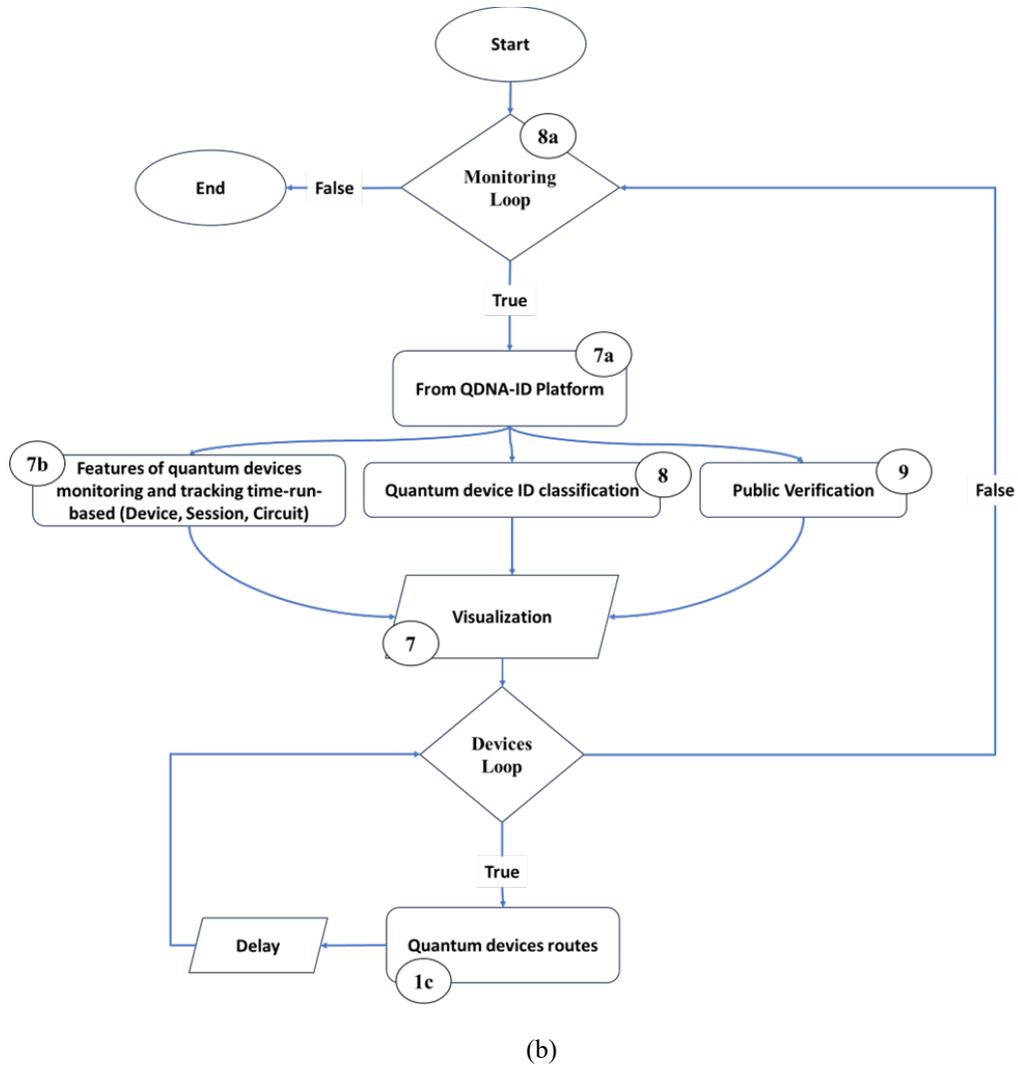

(b)

Figure 2. Flowchart QDNA-ID platform claim "A complete trust chain that connects physical quantum behavior to digitally verifiable"

### 3. Circuits and measurements

QDNA-ID uses different quantum circuits to test quantum devices, but all circuits perform the same function. QDNA-ID starts with a direct measurement of each qubit, without any circuitry measuring the readout bias. If the device is ideal, then p0 ≈ 1 for the qubit initialized to |0⟩. Its importance in the fingerprint is the constant baseline for each run, and its change indicates drift in the measurement calibration. The first circuit measures the 1-qubit gates, a finite random sequence of RX/RY/RZ rotations, followed by a measurement. This is important in demonstrating the fidelity of 1-qubit gates and the accumulation of noise with depth (coherent with stochastic). Its importance in the fingerprint is the rich output distribution, which increases the measurement of entropy and other statistical metrics, and their change over time reveals drift in 1-qubit gates [13], [14].

The second circuit is a Ramsey circuit: H → idle → RZ(ϕ) → H → T2 coherence measurement and low-freq noise cancellation. In the fingerprint, if the device is good or recently calibrated after maintenance, it will show a recovery of occupation compared to Ramsey, meaning lower entropy and p0 closer to one [15]. This difference between T2* and T2 carries a clear physical signature. The third circuit is interleaved gate stress, which means repeating the same gate X or H (or ID) consecutively, as it measures the systematic cumulative error (over/under-rotation, phase error) of a single gate. The importance of the interleaved gate stress circuit in the fingerprint is that it shows a constant pattern at p0 and perplexity that depends on the gate type and the number of iterations, meaning that the pattern changes over time equals the drift in the calibration of that gate. The second circuit is a two-qubit entangler chain, which executes H on everyone, followed by a CNOT chain in a random order (determined by the Seed), which measures the quality of the entanglement gates, along with the topology, connectivity, and two-qubit errors [16]. Cross-talk probability is a strong and variable rotation on the zero qubit while the rest of the qubits remain idle. It measures the crosstalk and its effect on neighboring qubits. If there is cross-talk in the fingerprint, you will see deviations on the silent qubits (entropy rise or p0 change), because a device with better hardware provides greater stability here [17]. CHSH with Bell circuits: Set up a Bell pair, then four measurement settings (A, B), (A, B′), (A′, B), (A′, B′) to calculate the correlation values $E$, then the CHSH as in (1):

$$S = E(AB) + E(AB') + E(A'B) - E(A'B'), \quad CHSH\ threshold \geq 2 \qquad (1)$$

CHSH with Bell circuits is the origin lock for QDNA-ID, proving that the source is a real quantum source and not a simulation, thus increasing the credibility of all subsequent fingerprints [18]. All these circuits serve multidimensional coverage. Small variations in X/H/CX calibrations, T2/T2* characteristics, and crosstalk create a stable "statistical signature" for each device, as well as durability over time: the presence of drift-sensitive elements allows for detection of variation and increases authentication confidence.

The strength of QDNA-ID Seed-based, which makes breaking circuit predictions computationally expensive, reducing the risk of deliberate forgery or tampering. The importance of the Seed is that it is a constant value in circuits that guarantees reproducibility. It allows the same circuit to be regenerated in the future to verify the fingerprint or to compare devices. Without a constant Seed, it is impossible to determine whether the change in results is due to the device or the circuit itself. The second advantage it offers is that the seed provides controlled diversity. When the Seed is deliberately changed, new variables of the same pattern are created (different randomness but with the same overall statistics). This diversity increases the richness of the quantitative fingerprint and prevents reliance on a single type of circuit. Fingerprint diversity is also one of the advantages of the Seed. Each device interacts with these random sequences differently (due to different calibrations and noise). The results of the features for each Seed form distinct points in the fingerprint space. The dispersion between seeds within the same device is less than the dispersion between devices, a property used for identity verification. Finally, security control over test generation revolves around storing the Seed within metadata, which allows for revalidation of the session without exposing the gateways themselves (can regenerate it later in a closed-source form). This prevents tampering with the results or claiming that different operations are the result of the same circuit.

4. Security mechanisms by cryptography

Security in cryptography involves encryption methods and digital signatures that include several types of encryptions, including symmetric, asymmetric, and hybrid [19]. Symmetric encryption uses the same key for encryption and decryption and is used for high confidentiality, such as AES Advanced Encryption Standard (AES), ChaCha20, characterized by very high speed with applications are file or database encryption. Asymmetric encryption uses two different keys (public and private) used for authentication with key exchange, such as RSA, ECC, and Ed25519, does not require a pre-shared secret key, but is slower than symmetric encryption and its applications are session keys or digital signatures [18]. Hybrid encryption uses asymmetric encryption to exchange the AES key and then symmetrically encrypt the actual data. It combines efficiency with comprehensive security, and its applications are in secure email encryption [20]. Hash functions are one-way mathematical functions that ensure data integrity, such as SHA-256, SHA3, and BLAKE2. Their properties are that no input can produce the same output (collision resistance). Any slight change produces a completely different hash. MAC or HMAC (Hash-based Message Authentication Code) are a secret key with a hash used to verify the source with integrity, i.e., prove that the sender knows the key (authenticity) and that the message has not been altered (integrity) [21]. The difference between a hash and an HMAC is that a hash only verifies the integrity of the data, while an HMAC verifies identity with integrity, such as HMAC-SHA256, CMAC (AES-based MAC) [22]. Digital Signatures such as public and private keys with a hash are used to prove the sender's identity (authenticity) with non-repudiation. Anyone can verify using the public key, such as RSA-SHA256, ECDSA, or Ed25519 [23]. Table 2. Illustrates the cryptography encryption and authentication.

Table 2. Summary of cryptography encryption and signature.

| Category | Keys | Primary Goal |
|---|---|---|
| **Symmetric Encryption** | AES, ChaCha20, DES | Confidentiality |
| **Asymmetric Encryption** | RSA, ECC, Ed25519 | Authentication or Key Exchange |
| **Hash Functions** | SHA-256, SHA3, BLAKE2 | Data Integrity |
| **Message Authentication Codes (MAC / HMAC)** | HMAC-SHA256 | Source Verification with Integrity |
| **Digital Signature** | RSA-SHA256, ECDSA, Ed25519 | Non-repudiation with Public Authenticity |
| **Hybrid Encryption** | RSA & AES | Efficiency and Overall Security |

Integrity, authenticity non-repudiation, provenance, and timing are used in QDNA-ID. QDNA-ID starts with a seed in quantum circuits to prevent circuit breakers, with CHSH attestation to separate real devices from emulators, and internal integrity and authentication is represented by a 256-bit (32-byte) symmetric HMAC key with the HMAC-SHA256 algorithm. For authenticity non-repudiation, it uses RSA private and public RSA-SHA256 algorithms with an asymmetric key size of 2048 bits (upgradable to 3072+). Provenance copyright is included in the digital signature with institutional proof of ownership within the message. Table 2. Illustrates the cryptography keys used in QDNA-ID.

Table 3. QDNA-ID cryptography keys used for integrity and provenance timing-based.

| Security Layer | Technology | Purpose |
| --- | --- | --- |
| **HMAC-SHA256** | Internal Secret Check | Detects any internal tampering with the file |
| **RSA-SHA256** | Public Digital Signature | Proves the file is from the original source |
| **SHA256** | Feature Hashing | Generates a stable digital fingerprint |
| **CHSH attestation** | Physical Test | Proves the measurement is from a real quantum device, not a simulator |

## 5. Features extraction and classification

Feature extraction is the third stage after digitally signing the results of circuit injections into quantum devices and extracting the results. The QDNA-ID platform uses Probability Distribution and Shannon Metrics (Shannon Entropy, Normalized Entropy, Gini Impurity, Effective Support / Perplexity). Entropy is a measure of the quantum randomness in the output. High entropy means high values, while low entropy means confined results or faint noise. It is used as an indicator of drift when coherence deteriorates over time, as in (2).

$$H(p) = - \sum_{i=1}^{K} p_i \ln p_i \tag{2}$$

The support measure, a measure of the number of possible states with non-zero probability, shows the actual diversity of the device's outputs. If support is low, the device shrinks to a small number of probabilities. If support is fully quantum, the diversity of states is maintained. The normalization measure of entropy relative to the support size makes it possible to fairly compare circuits of different sizes (2 qubits ≠ 4 qubits) with a range between 0 (completely ordered) and 1 (completely random), as in (3).

$$H_{norm}(p) = \frac{H(p)}{\ln K} \tag{3}$$

Perplexity is a measure of the number of "effective" states over which a system distributes its probability. It is used in linguistics and artificial intelligence. In QDNA-ID, it indicates the number of "true" states the quantum system activates. A high value indicates a wide superposition, while a low value indicates a narrow output, as in (4).

$$Perplexity\ (p) = \exp(H(p)); \qquad Perplexity \in [1, K] \qquad (4)$$

The Gini measure is a measure of the inequality of the P distribution. The closer it is to 1, the more balanced and widespread it is. The closer it is to 0, the more concentrated the distribution is in a few states, as in (4). A measure attached to the efficiency appendix, i.e., the number of effective states after Gini correction or an estimate of the true number of states participating in the superposition.

$$Gini(p) = 1 - \sum_{I=1}^{K} p_i^2; \qquad (5)$$

Parity and Bit-Level Features also used in QDAN-ID such as (Parity Bias, Bernoulli Variance). Parity Bias (measurement bias) compares even vs odd bit-parity outcomes considered sensitive to systematic noise or readout bias as in (6):

$$Parity = P_{EVEN} - P_{ODD} = 2P_{EVEN} - 1 \qquad (6)$$

Bernoulli Variance describes the "sharpness" or "flattening" of the output distribution, with a large variance indicating that the device produces preferred (non-equilibrium) states. A small variance indicates that the results are nearly homogeneous (or highly noisy). In short, variance is the variance of a probability distribution as in (7):

$$Var = p_0(1 - p_0) \qquad (7)$$

For Distance from Uniformity QDNA-ID used Total Variation Distance (TV), Kullback Leibler Divergence (KL), Jensen Shannon Divergence (JS). To test how "physical" or biased the quantum device behavior QDNA-ID compare $p$ to a uniform distribution $u$ by the TV as in (8):

$$TV\ (p, u) = \frac{1}{2} \sum_i |p_i - u_i| \qquad (8)$$

Where; 0 means perfectly uniform and 1 means fully deterministic for TV. The KL measures how much information separates $p$ from uniform $u$ as in (9):

$$KL(p||u) = \sum_i p_i \ln \frac{p_i}{u_i} = \ln K - H_p \qquad (9)$$

The JS bounded metric of similarity used to smooth measures the distance between the actual distribution of the device and a uniform distribution, showing the extent to which the device deviates from "ideal", unbiased behavior. A small value indicates nearly balanced output (good superposition), while a large value indicates a clear deviation or noise pattern as in (10):

$$JS(p||u) = \frac{1}{2} KL\left(p || \frac{p+u}{2}\right) + \frac{1}{2} KL\left(u || \frac{p+u}{2}\right) \qquad (10)$$

When the same circuits are executed by multiple sessions QDNA-ID tracks changes of drift in measured entropy by Temporal Drift Across Runs (Drift Index, Session Distance Matrix). Drift Index quantifies accumulated change in device behavior as in (11):

$$Drift\ (t) = \sum_c |m_c^t - m_c^{t-1}| \tag{11}$$

Where $m_c^t$ is the feature of circuit $c$ at time $t$. Session Distance Matrix measures how far apart two experimental sessions are in feature space as in (12):

$$D_{ij} = \sum_c \left| m_c^{(i)} - m_c^{(j)} \right| \tag{12}$$

To check how different features interact, Correlation analysis is used in QDNA-ID helps identify which quantum metrics evolve together. Table 3. Illustrates the features measurements for QDNA-ID drifts-time-based. Table 4. Illustrates the features metric used in QDNA-ID platform.

Table 4. QDNA-ID features measurements for drifts-time-based.

| Metric | Describes | Significance |
| --- | --- | --- |
| $p_0 \backslash p_1$ | Measurement bias | SPAM calibration |
| Variance | Sharpness of distribution | Balance vs. centralization |
| Entropy | Resulting randomness | Overlap and noise |
| Perplexity | Number of effective states | Output complexity |
| Gini | Energy distribution | State balance |
| Jensen–Shannon divergence | Distance from ideal distribution | Device fingerprint metric |
| Drift Index | Session change magnitude | Device aging monitoring |

Data processing and feature extraction based on statistical feature files extracted from the difference measurement of the intersecting quantum circuits. Using the intersection between and relying on the intersection of the quantum circuits reduces missing data and improves comparability. The features are converted into a fixed-length numerical vector, forming a structured matrix suitable for machine learning. Three machine learning pipelines were used to classify quantum devices:

- **Nearest Centroid with L1 (Manhattan) distance:** Provides simplicity and robustness against outliers by using a baseline reference that computes the centroid of each class and assigns new samples based on minimum L1 distance [24].
- **Logistic Regression with L2 regularization:** Provides interpretable probabilistic linear terms through an optional class weighting mechanism to address class imbalance with L2 regularization to control overfitting [25].

- **Random Forest:** Provides a robust nonlinear baseline for structured circuit data by using an ensemble of decision trees for nonlinear dependencies and variable interactions that linear models cannot capture [26].

The machine learning algorithms were implemented using a cross-validation strategy with permutation testing to increase the robustness and reliability of the results (See Figure 3.).

```
Input: devA, devB, metrics, min_presence, CV params, model params

# Data preparation
filesA, filesB ← list_feature_files(devA), list_feature_files(devB)
CIRCUITS ← circuits present in ≥(min_presence·N) of both devices
for each file in (A∪B):
    x ← [feat[c][m] for c∈CIRCUITS, m∈metrics]
    X.append(x); y.append(label)

# Models
MODELS = {
  NC: Imputer→RobustScaler→NearestCentroid(L1),
  LR: Imputer→StandardScaler→LogReg(L2),
  RF: Imputer→RandomForest(depth, leaf)
}

# Cross-validation
for model in MODELS:
  for (train,test) in StratifiedKFold(X,y):
      fit(model, X[train], y[train])
      scores ← model.predict_proba/test
      if opt_threshold: thr←best_threshold(y[test],scores)
      ŷ ← (scores≥thr)
      record acc, auc, F1, CM

# Permutation test
true_acc ← mean_CV_accuracy(NC,X,y)
pval ← (1 + #perms with shuffled-y acc ≥ true_acc) / (1 + N_perm)

# Output
save metrics, predictions, confusion matrices, permutation pval
```

Figure 3. Pseudocode of machine learning classification algorithms.

## 6. Experimental Results

The QDNA-ID platform for real quantum devices was developed using several statistical metrics to accurately measure quantum identity. The platform collected 25 sessions with a constant 1024 shots for all sessions from 12/10/2025 to 20/10/2025 on two real IBM quantum computer, ibm_torino and ibm_brisbane, approximately every 8 hours. The QDNA-ID platform uses statistical metrics to present the differences between all intersecting circuits between sessions, from session to session, and from computer to computer, with real-time provenance tracking and monitoring. Sessions and features for each quantum computer are authenticated and physically protected with a CHSH quantum circuit to separate real machines from simulated or tampered ones, as well as HMAC-SHA256 keys to protect against internal tampering, and RSA-SHA256 public and secret encryption to verify sessions for public use for increased reliability. To differentiate between devices using statistical features, sessions, and circuits, mean, median, and median-bias were used for feature metrics. Median-bias is a statistical algorithm used to reduce the influence of outlier values. The average uncertainty used to measure randomness for both devices ranges between 0.22 and 0.26. The Torino tends to have higher peaks and more frequent values than Brisbane, indicating a greater probability range for Torino. The Brisbane exhibits sharper swings between sessions 3-8 suggesting it is more sensitive to noise. The similarity of values in session 13 suggests that they are influenced by a single circumstance related to similar updates or calibrations. ibm-torino's randomness metric provides a more reliable and mature quantitative identity over time (See Figure 4).

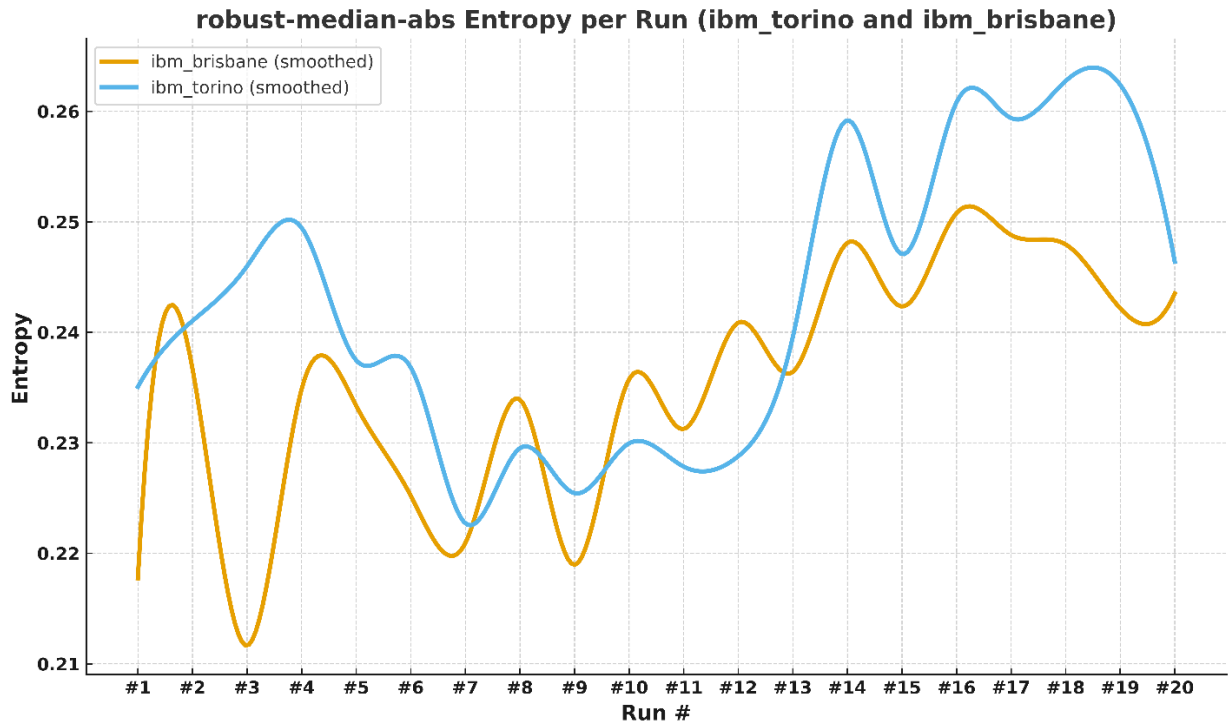

Figure 4. Median-bias Entropy per run for each quantum device.

The Gini metric shows higher values for the Torino and then gradually decreases compared to the Brisbane. A high Gini means the normal distribution is non-homogeneous due to noise in the qubits. The Torino starts with a clear bias and then gradually improves, perhaps due to recalibration or thermal stabilization. In contrast, the Brisbane starts with a static noise condition and then becomes more biased over time, indicating either coherence degreasing or thermal drift. The two devices exhibit opposite behavior, one moving toward stability and the other toward drift

(See Figure 5a). The JS divergence metric measures the distance between the actual distribution of results and the ideal uniform distribution. The higher values for the Brisbane device indicate that it is further from ideal randomness. The Torino device shows more stable values between 0.04-0.06, meaning it is closer to a real random distribution. The synchronized fluctuations between the two devices suggest a common update or environmental influence affecting both. The Torino device exhibits higher probabilistic stability compared to the Brisbane device, which is more susceptible to fluctuations (See Figure 5b). The complexity measure quantifies the number of effective states in the quantum distribution (the degree of randomness). The Turin device starts with a high complexity of 0.39 and then gradually decreases to 0.33, while the Brisbane device starts low and then gradually increases, stabilizing at 0.36. The Turin device loses diversity over time, while the Brisbane device temporarily improves before stabilizing (See Figure 5c). The differences in these measures confirm that each device possesses a unique quantum signature that evolves independently in real time.

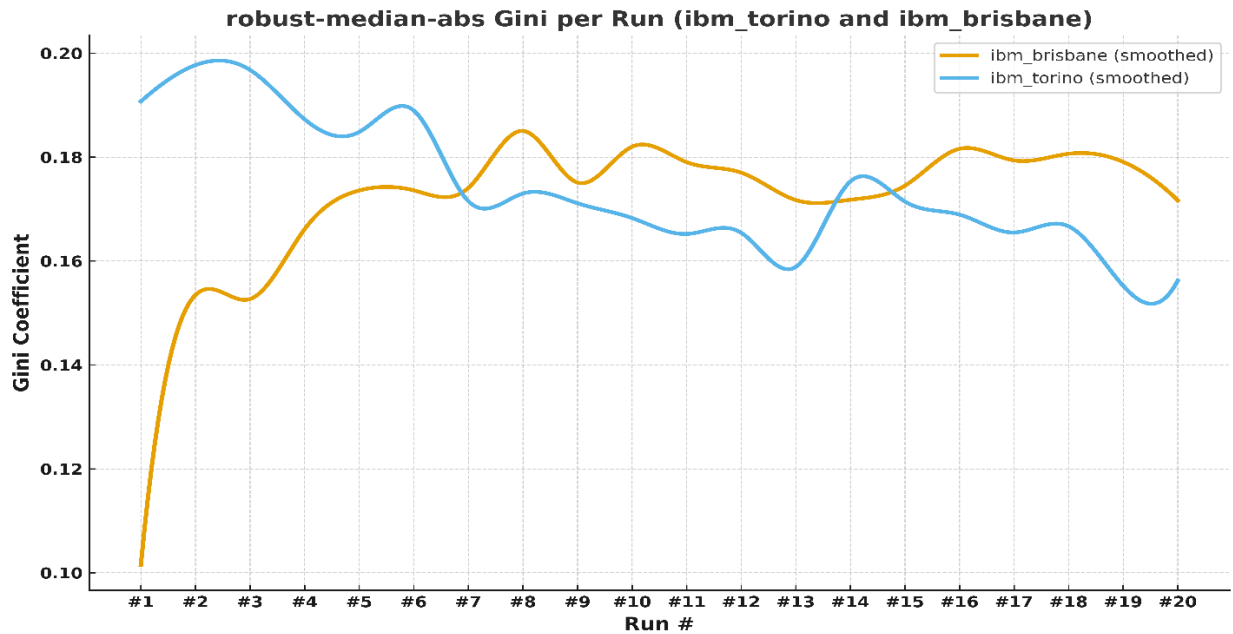

(a)

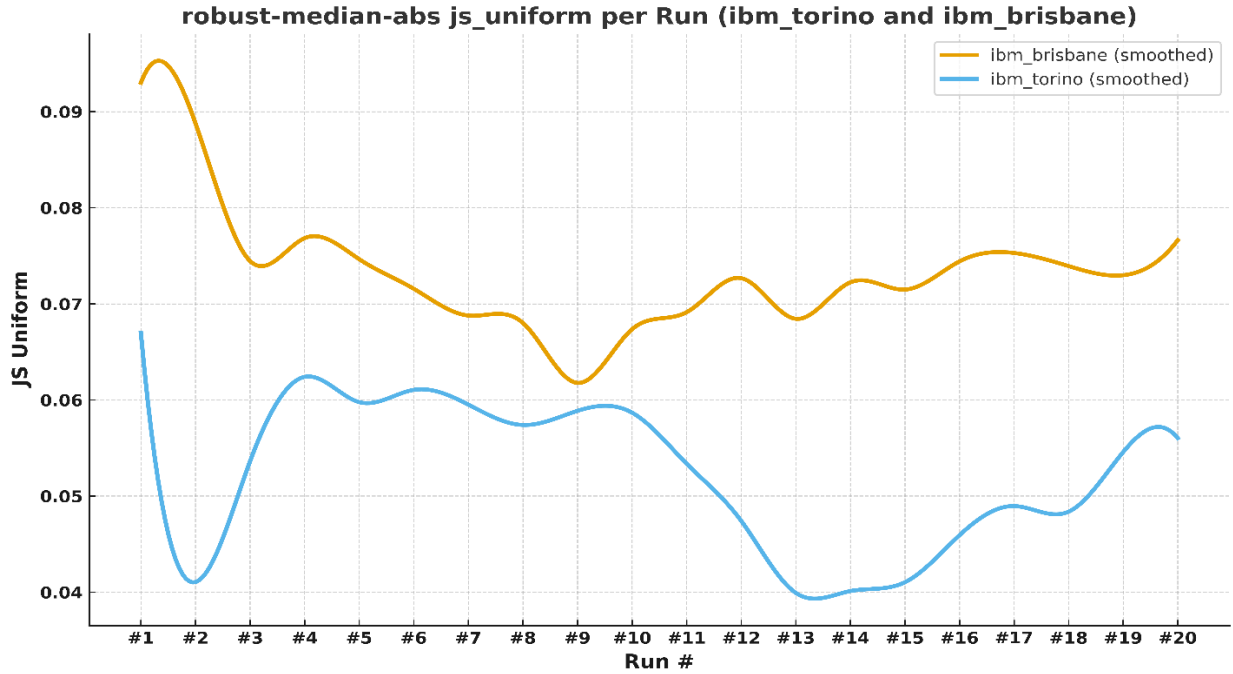

(b)

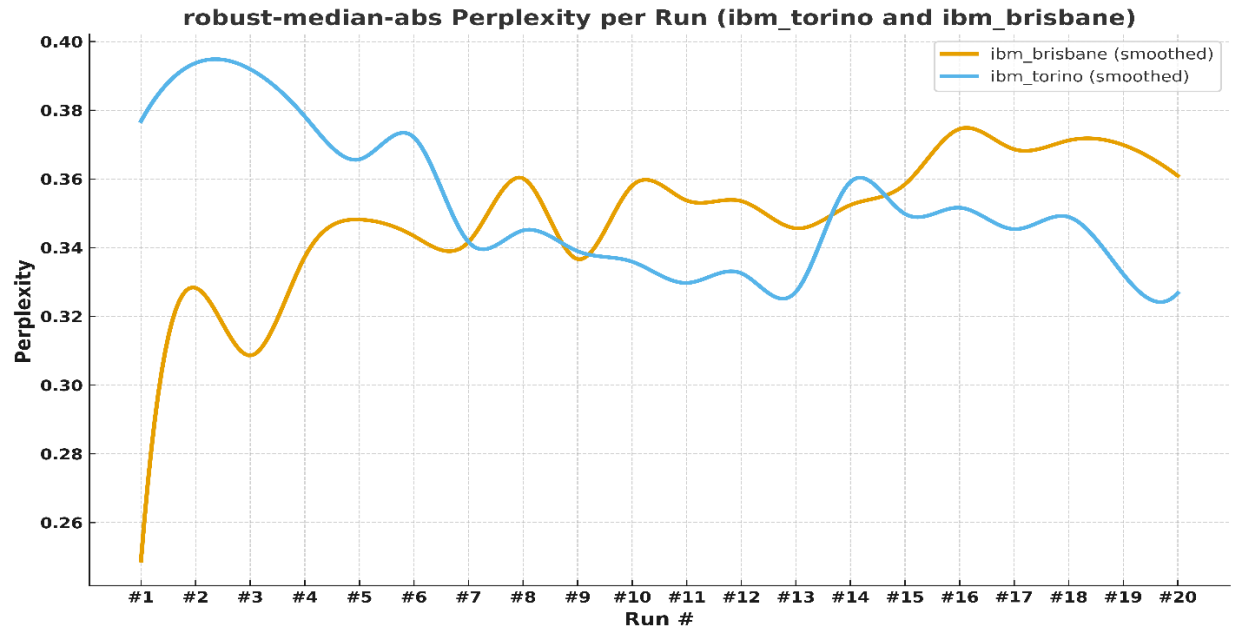

(c)

Figure 5. Median-bias features (Gini, Jensen–Shannon, Perplexity) per run for each quantum device.

The total drift index over time shows continuous physical changes in the quantum devices due to recurring peaks. Brisbane shows higher and wider peaks than Torino, while the temporal correlation between the two devices in the middle of the period indicates a calibration or update of both devices. Torino device in the latter half of the measurements, shows greater stability than

Brisbane, meaning it regained its physical equilibrium over time, while Brisbane continued to experience significant drift (See Figure 6).

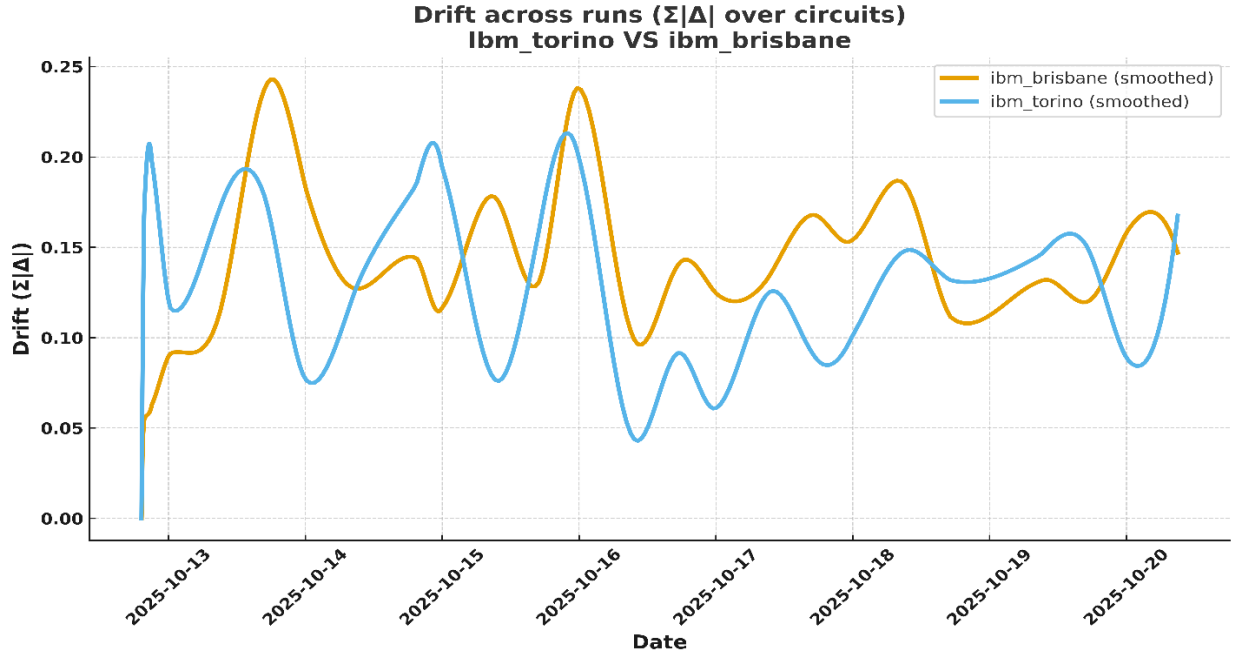

Figure 6. Total summation of drift per run for each quantum device

After the results of 25 experimental sessions for each device, the Torino device showed more stability and consistency in its quantum characteristics than the Brisbane device, which showed noticeable fluctuations and drifts. The differences in statistical features for each device confirms the unique of quantum identity (See Table 5.).

Table 5. Summary of QDNA-ID real-time quantum devices monitoring.

| Metric | Torino | Brisbane | Interpretation |
|---|---|---|---|
| **Entropy** | High and stable | Lower and variable | Torino is richer in outcomes |
| **Gini** | Gradually decreases | Gradually increases | Torino stabilizes. Brisbane deteriorates |
| **JS Divergence** | Low and stable | High and fluctuating | Torino is closer to ideal randomness |
| **Perplexity** | Gradually decreases | Gradually increases | Exchange in physical state |
| **Drift (ΣΔ)** | Gradually decreases | Retains fluctuations | Torino is physically stable |

The basis for measuring overall randomness is built on the intersecting quantum circuits between the 8 quantum devices, with 4 different circuits per device, for a total of 12 circuits per session. The QDAN-ID platform not only tracks the overall temporal behavior of the devices but also provides detailed information on the characteristics of each circuit and each session. The Torino device in sessions 24 and 25 shows an entropy drift with a total value of 2.6634, with no change in p0=0.000, meaning the state distribution remained unchanged. The largest contributions to the entropy drift were in the spin-echo, pm, and interleaved_x circuits, which represent the phase stability and coherence (T2) circuits of the qubits. The Torino device shows high sensitivity but less temporal stability indicates that it is susceptible to thermal and frequency fluctuations in real-world quantum systems (See Figure 7a). The Brisbane device showed less drift when comparing sessions 24 and 25, with a total value of 0.9239, indicating better stability and performance consistency. The spin-echo and interleaved_x circuits showed the most drift, along with the crosstalk circuit, which contributed only slightly indicates a minor fluctuation in the cross-coupling between the qubits without significant loss of coherence (See Figure 7b). The Brisbane device shows ideal quantum verification (quantum provenance and reproducibility) for experiments, measurements, and comparisons.

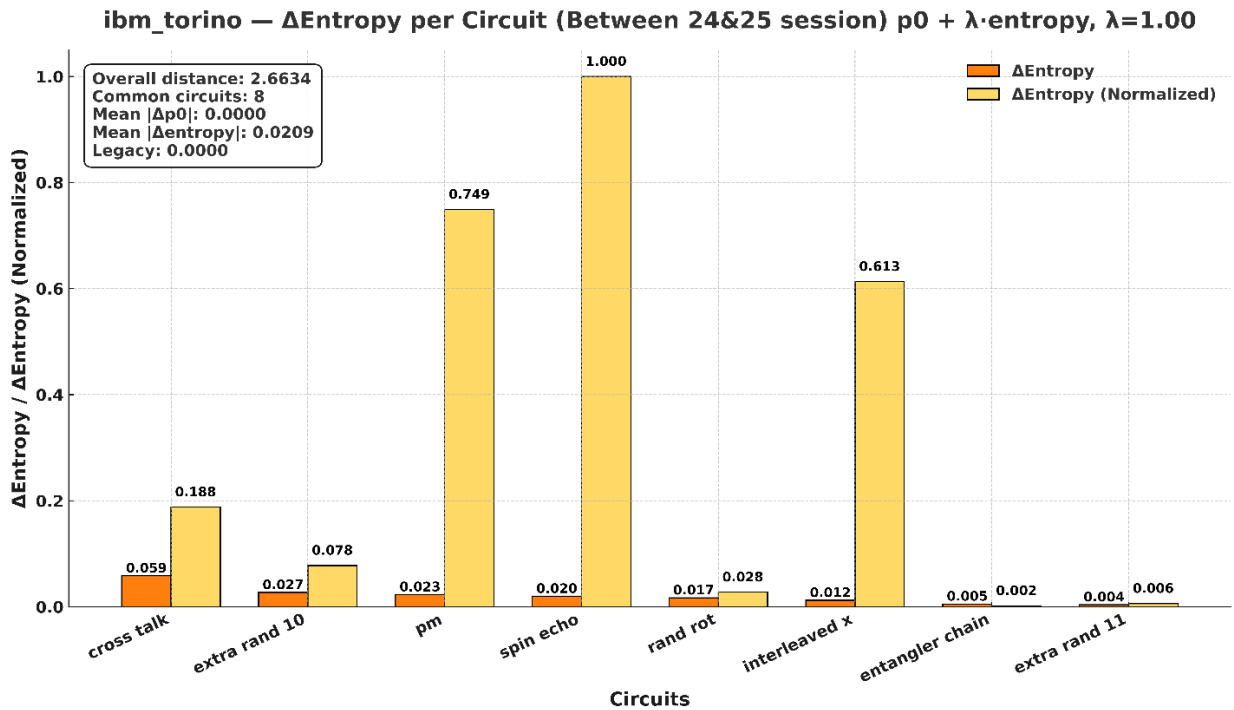

(a)

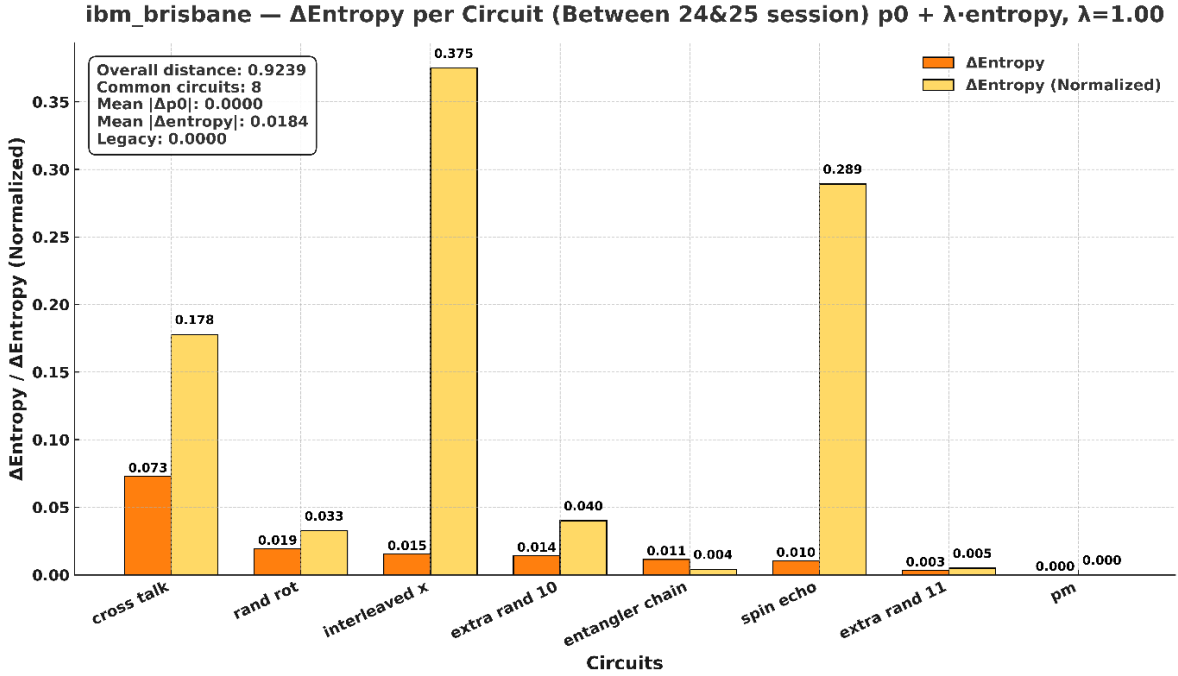

(b)

Figure 7. Delta-Entropy per circuit for 24 & 25 session: a) Torino device. b) Brisbane device.

The classification results between the two devices using machine learning were obtained using three algorithms: Nearest Centroid, Random Forest, and Logistic Regression. The features used were QDNA-ID 'p0', 'entropy', 'js_uniform', 'gini', and 'parity_bias' between the two quantum devices using a cross-validation strategy with 6 folds to ensure reliability due to the limited number of sessions per device. The machine learning algorithms showed fluctuating classification results in the early sessions until they stabilized after 20 sessions for each device. The best overall performance was achieved by the Nearest Centroid algorithm with an accuracy of 94.21% ± 0.09 and a p-value of 0.001 from the permutation test. The high performance achieved by this algorithm without using class weighting or threshold optimization is strong evidence of the unique identity of the quantum devices.

The Random Forest algorithm demonstrated the model's strong ability to distinguish non-linear relationships between the metrics with an accuracy of 90.28% and an AUC of 97%. The relatively high AUC indicates the model's high sensitivity in distinguishing between the devices across all possible threshold points. The Logistic Regression algorithm performed the worst of the three models, but it was relatively stable with an accuracy of 88.18% and an F1-score of 87.5%. The Logistic Regression results show that linear separation is only partially possible. The classification algorithm results show the robustness of the QDNA-ID trust chain in separating and distinguishing quantum device identities (See Table 6.).

Table 6. Confusion reports metric for classification the quantum devices.

| Model | Accuracy (Mean) | AUC (Mean) | AUC (Std) | Precision (Macro) | Recall (Macro) | F1 (Macro) | Permutation p-value |
|---|---|---|---|---|---|---|---|
| **Nearest Centroid L1** | **0.9421** | 0.9375 | 0.0955 | **0.9595** | 0.9375 | 0.9372 | **0.0010** |
| **Random Forest (baseline)** | 0.9028 | **0.9708** | 0.0466 | 0.9354 | 0.8958 | 0.8884 | – |
| **Logistic Regression (L2)** | 0.8819 | 0.9083 | 0.0796 | 0.9123 | 0.8750 | 0.8754 | – |

The tests and methodology were conducted on only two real quantum devices with a relatively short duration and limited number of sessions, so generalization is one of the limitations of QDNA-ID. The operational cost and computational cost of sessions due to the high queue on the machines are among the most significant limitations. The CHSH attestation test is not theoretically impossible to falsify if an attacker has access to a real device or real data, but it increases the cost of spoofing. Signatures protect integrity but do not prevent leakage if the keys are disclosed, so separate keys and a revocation plan are required. QDNA-ID uses only 1024 shots in quantum circuits, which also makes it limited, but 1024 shots is better than 256 and 512 shots, which is considered better in terms of noise. Results may vary based on provider updates or daily calibrations of quantum computer providers.

## 7. Conclusion

QDNA-ID is a provenance system and verification platform for quantum devices that links a device's true physical behavior to cryptographically verifiable digital records. The goal of the system is for each quantum device operation to generate a unique physical quantum fingerprint time-based that can be subsequently tracked and verified. QDNA-ID collects quantum device (QDNA-ID uses ibm_torino & ibm_brisbane) operation data at three times a day, every 8 hours, by injecting more than 10 quantum circuits (QDNA-ID uses 12) into the same job within the quantum computer, with different shots ranging from 256 to 2048 (QDNA-ID is set to 1024 for all devices). The raw data is then converted into numerical features (QDNA-ID uses entropy, Gini, Jensen Shannon divergence, perplexity, and p0). The uniqueness of QDNA-ID is its Signature and Timestamping with provenance record, which provides a quantum-computer-based-trust-chain concept by creating a file containing the fingerprint, HMAC, RSA signature, and provenance data, linking the fingerprint to a trusted time and source.

## References

[1] J. Wu, T. Hu, and Q. Li, "Detecting Fraudulent Services on Quantum Cloud Platforms via Dynamic Fingerprinting," in Proceedings of the 43rd IEEE/ACM International Conference on Computer-Aided Design, 2024, pp. 1–8.

[2] V. Mutolo et al., "Quantum Computer Fingerprinting using Error Syndromes," arXiv preprint arXiv:2506.16614, 2025.

[3] T. Li, Z. Zhao, and J. Yin, "Task-driven quantum device fingerprint identification via modeling qnn outcome shift induced by quantum noise," in Companion Proceedings of the ACM Web Conference 2024, 2024, pp. 557–560.

[4] Y. Wang and E. S. Ismail, "A Review on the Advances, Applications, and Future Prospects of Post-Quantum Cryptography in Blockchain, IoT," IEEE Access, 2025.

[5] S. E. Bootsma and M. De Vries, "A Survey on the Quantum Security of Block Cipher-Based Cryptography," IEEE Access, 2024.

[6] S. Cherbal, A. Zier, S. Hebal, L. Louail, and B. Annane, "Security in internet of things: a review on approaches based on blockchain, machine learning, cryptography, and quantum computing," J Supercomput, vol. 80, no. 3, pp. 3738–3816, 2024.

[7] J. Wu, T. Hu, and Q. Li, "Q-id: Lightweight quantum network server identification through fingerprinting," IEEE Netw, vol. 38, no. 5, pp. 146–152, 2024.

[8] A. Shokry and M. Youssef, "QLoc: A realistic quantum fingerprint-based algorithm for large scale localization," in 2022 IEEE International Conference on Quantum Computing and Engineering (QCE), IEEE, 2022, pp. 238–246.

[9] S. Martina, L. Buffoni, S. Gherardini, and F. Caruso, "Learning the noise fingerprint of quantum devices," Quantum Mach Intell, vol. 4, no. 1, p. 8, 2022.

[10] G. Vrana, D. Lou, and R. Kuang, "Raw qpp-rng randomness via system jitter across platforms: a nist sp 800-90b evaluation," Sci Rep, vol. 15, no. 1, p. 27718, 2025.

[11] K. Mansoor, M. Afzal, W. Iqbal, and Y. Abbas, "Securing the future: exploring post-quantum cryptography for authentication and user privacy in IoT devices," Cluster Comput, vol. 28, no. 2, p. 93, 2025.

[12] R. Arnon-Friedman, F. Dupuis, O. Fawzi, R. Renner, and T. Vidick, "Practical device-independent quantum cryptography via entropy accumulation," Nat Commun, vol. 9, no. 1, p. 459, 2018.

[13] R. Jozsa, "On the simulation of quantum circuits," arXiv preprint quant-ph/0603163, 2006.

[14] M. P. A. Fisher, V. Khemani, A. Nahum, and S. Vijay, "Random quantum circuits," Annu Rev Condens Matter Phys, vol. 14, no. 1, pp. 335–379, 2023.

[15] R. Kaubruegger, D. V Vasilyev, M. Schulte, K. Hammerer, and P. Zoller, "Quantum variational optimization of Ramsey interferometry and atomic clocks," Phys Rev X, vol. 11, no. 4, p. 041045, 2021.

[16] H. Terashima and M. Ueda, "Nonunitary quantum circuit," International Journal of Quantum Information, vol. 3, no. 04, pp. 633–647, 2005.

[17] X. Dai et al., "Calibration of flux crosstalk in large-scale flux-tunable superconducting quantum circuits," PRX Quantum, vol. 2, no. 4, p. 040313, 2021.

[18] M. Dušek, N. Lütkenhaus, and M. Hendrych, "Quantum cryptography," Progress in optics, vol. 49, pp. 381–454, 2006.

[19] C. Paar, J. Pelzl, and T. Güneysu, "Introduction to public-key cryptography," in Understanding Cryptography: From Established Symmetric and Asymmetric Ciphers to Post-Quantum Algorithms, Springer, 2024, pp. 177–203.

[20] S. Ricci, P. Dobias, L. Malina, J. Hajny, and P. Jedlicka, "Hybrid keys in practice: Combining classical, quantum and post-quantum cryptography," IEEE Access, vol. 12, pp. 23206–23219, 2024.

[21] G. Brassard, P. Høyer, and A. Tapp, "Quantum cryptanalysis of hash and claw-free functions," in Latin American Symposium on Theoretical Informatics, Springer, 1998, pp. 163–169.

[22] A. Soni, S. K. Sahay, and V. Soni, "Hash Based Message Authentication Code Performance with Different Secure Hash Functions," in 2025 10th International Conference on Signal Processing and Communication (ICSC), IEEE, 2025, pp. 104–109.

[23] V. Mavroeidis, K. Vishi, M. D. Zych, and A. Jøsang, "The impact of quantum computing on present cryptography," arXiv preprint arXiv:1804.00200, 2018.

[24] R. Suwanda, Z. Syahputra, and E. M. Zamzami, "Analysis of euclidean distance and manhattan distance in the K-means algorithm for variations number of centroid K," in Journal of Physics: Conference Series, IOP Publishing, 2020, p. 012058.

[25] A. B. Musa, "Comparative study on classification performance between support vector machine and logistic regression," International Journal of Machine Learning and Cybernetics, vol. 4, no. 1, pp. 13–24, 2013.

[26] Y. Liu, Y. Wang, and J. Zhang, "New machine learning algorithm: Random forest," in International conference on information computing and applications, Springer, 2012, pp. 246–252.